\documentclass[journal]{IEEEtran}

\ifCLASSINFOpdf
\else
   \usepackage[dvips]{graphicx}
\fi
\usepackage{url}

\usepackage{graphicx}
\usepackage{amsmath}
\usepackage{amsfonts}
\usepackage{bm}
\usepackage{algorithm,algorithmic}
\usepackage{mathtools}
\usepackage{comment}
\usepackage{cite}

\DeclareMathOperator*{\minimize}{minimize}
\DeclareMathOperator*{\argmin}{arg~min}
\DeclareMathOperator{\sign}{sign}
\DeclareMathOperator{\prox}{prox}

\newcommand{\abs}[1]{\left\lvert#1\right\rvert}
\newcommand{\norm}[1]{\left\lVert#1\right\rVert}

\newcommand{\paren}[1]{\left(#1\right)}
\newcommand{\sqb}[1]{\left[#1\right]}

\newcommand{\curbra}[1]{\left\{#1\right\}}
\begin{document}

\title{Reconstruction of Piecewise-Constant Sparse Signals for Modulo Sampling}

\author{Haruka Kobayashi and Ryo Hayakawa, \IEEEmembership{Member, IEEE}
\thanks{
    This work will be submitted to the IEEE for possible publication. Copyright may be transferred without notice, after which this version may no longer be accessible.
}
\thanks{
    % Manuscript received April XX, 20XX; revised September XX, 20XX. 
    This work was supported in part by The Telecommunications Advancement Foundation and Japan Society for the Promotion of Science (JSPS) KAKENHI under Grant JP24K17277.
}%
\thanks{
    Haruka Kobayashi is with the Graduate School of Engineering Science, The University of Osaka, 560-8531, Osaka, Japan.
}%
\thanks{
    Ryo Hayakawa is with the Institute of Engineering, Tokyo University of Agriculture and Technology, 184-8588, Tokyo, Japan.
}}

\markboth{Preprint}
{Kobayashi \MakeLowercase{\textit{et al.}}: Reconstruction of Piecewise-Constant Sparse Signals for Modulo Sampling}
\maketitle

\begin{abstract}
Modulo sampling is a promising technology to preserve amplitude information that exceeds the observable range of analog‑to‑digital converters during the digitization of analog signals. 
Since conventional methods typically reconstruct the original signal by estimating the differences of the residual signal and computing their cumulative sum, each estimation error inevitably propagates through subsequent time samples. 
In this paper, to eliminate this error‑propagation problem, we propose an algorithm that reconstructs the residual signal directly. 
The proposed method takes advantage of the high‑frequency characteristics of the modulo samples and the sparsity of both the residual signal and its difference. 
Simulation results show that the proposed method reconstructs the original signal more accurately than a conventional method based on the differences of the residual signal.
\end{abstract}

\begin{IEEEkeywords}
Modulo sampling, dynamic range, unlimited sampling, ADMM.
\end{IEEEkeywords}

\IEEEpeerreviewmaketitle

\section{Introduction}

\IEEEPARstart{S}{ampling} plays a fundamental role in modern signal processing. 
An analog‑to‑digital converter (ADC) for sampling is typically characterized by two primary performance metrics: the sampling period and the observable amplitude range. 
When the maximum frequency of the input signal exceeds half of the sampling frequency, aliasing distortion arises during reconstruction~\cite{SamplingTheoryBandlimited-Eldar-2015,SubNyquistSampling-Mishali-2011}. 
Furthermore, when the signal amplitude exceeds the observable range, the signal is clipped and accurate recovery of the original signal becomes generally impossible.
Signal clipping can cause serious problems in a wide range of practical applications~\cite{VisionSensorHaving-Yamada-1998, DetectionReconstructionClipped-Bie-2015}.
To prevent such distortions, a higher sampling rate or a wider dynamic range is required, which in turn leads to increased power consumption. 
To balance energy efficiency and reconstruction accuracy, it is desirable to use an ADC with a lower sampling rate and a narrower dynamic range without causing  signal clipping.

\begin{figure}[tb]
  \centering
  \includegraphics[width=0.48\textwidth]{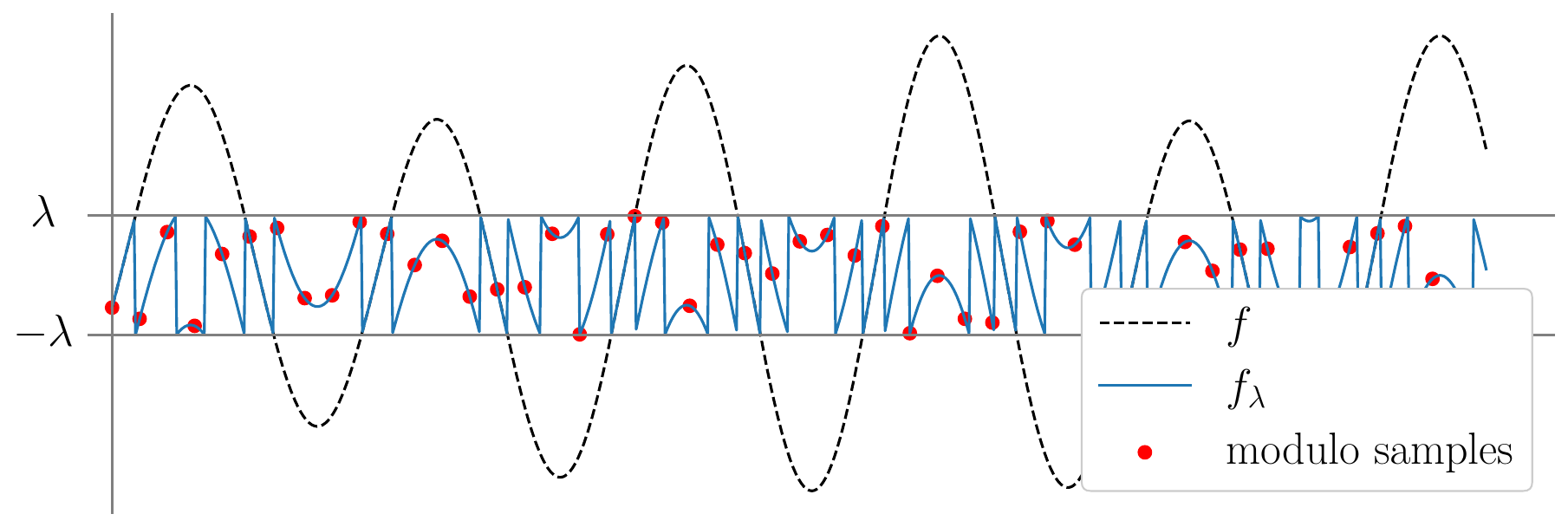}
  \caption{Illustration of modulo sampling.}
  \label{fig:modulo}
\end{figure}

To avoid clipping in an ADC with a small observable amplitude range, modulo sampling has been proposed~\cite{UnlimitedSamplingReconstruction-Bhandari-2021}.
In modulo sampling, a folding operation is performed by a modulo operator before measurement so that the amplitude of the resulting folded signal lies within $[-\lambda, \lambda)$ ($\lambda>0$).
The folded signal is then sampled by an ADC that can measure only the range $[-\lambda, \lambda)$.
Fig.~\ref{fig:modulo} illustrates the concept of modulo sampling.  
By comparing the original signal $f$ with the folded signal $f_\lambda$ obtained from modulo sampling, we can see that modulo sampling prevents clipping and preserves the waveform in part even when the amplitude exceeds the observable range.
However, since we sample the folded signal in modulo sampling, it is necessary to reconstruct the original unfolded signal from these samples.

Various methods have been proposed to reconstruct the original signal from the samples obtained by modulo sampling~\cite{WaveletBasedReconstructionUnlimited-Rudresh-2018,UnlimitedSamplingSparse-Bhandari-2018,IdentifiabilitySparseVectors-Prasanna-2021,ImplantableCMOSImage-Sasagawa-2016,UnlimitedSampling-Bhandari-2017, UnlimitedSamplingReconstruction-Bhandari-2021,NyquistRateModulo-Romanov-2019,ResidualRecoveryAlgorithm-Azar-2022,LassoBasedFastResidual-Shah-2023,UnlimitedSamplingTheory-Bhandari-2022,UnlimitedSamplingHysteresis-Florescu-2021,UnlimitedSamplingGeneralized-Florescu-2022}.
A hardware‑oriented approach~\cite{ImplantableCMOSImage-Sasagawa-2016} uses a reset count map to record the number of foldings together with the folded signal. 
This approach, however, requires complex electronic circuitry as well as additional power and memory.
As a reconstruction method without the reset count map, an algorithm using high‑order differences of the signal has been studied~\cite{UnlimitedSampling-Bhandari-2017, UnlimitedSamplingReconstruction-Bhandari-2021}.
Subsequently, a prediction-based algorithm has been proposed, demonstrating that in the absence of noise, perfect recovery is possible for finite energy signals at any sampling rate above the Nyquist rate~\cite{NyquistRateModulo-Romanov-2019}.
More recently, a more noise-robust method called beyond bandwidth residual reconstruction ($B^2R^2$), which focuses on the high-frequency band of the folded signal, has been proposed~\cite{ResidualRecoveryAlgorithm-Azar-2022}. Building on this approach, least absolute shrinkage and selection operator-$B^2R^2$ (LASSO-$B^2R^2$) has been introduced to further improve performance~\cite{LassoBasedFastResidual-Shah-2023}. 
This method exploits the sparsity of the first-order difference of the residual signal, which represents the number of wraps and is defined as the difference between the signal obtained by modulo sampling and the original signal.
Although this approach can reconstruct the signal efficiently, reconstruction methods based on the difference of the residual signal suffer from the problem that a reconstruction error at one time index propagates to all subsequent points.

In this study, we propose an optimization problem for directly reconstructing the residual signal in modulo sampling.
In the proposed residual signal-based optimization problem, we first introduce a regularization term to exploit the sparsity of the first-order difference of the residual signal as in conventional methods.
In addition, we newly focus on the sparsity of the residual signal itself and introduce a regularization term that promotes this property as well.
To solve the optimization problem efficiently, we derive an algorithm based on the alternating direction method of multipliers (ADMM)~\cite{Eckstein1992-ra,DistributedOptimizationStatistical-Boyd-2010}.
Simulation results show that the proposed method achieves a lower normalized mean squared error (NMSE) than the conventional LASSO‑$B^2R^2$.

Throughout this paper, $\mathbb{R}$, $\mathbb{C}$, and $\mathbb{Z}$ denote the sets of all real numbers, complex numbers, and integers, respectively.
For a vector $\bm{x} = \sqb{x_{1}\ \dotsb\ x_{N}}^{\top} \in \mathbb{R}^{N}$, $\norm{\bm{x}}_{p} = \sqrt[p]{\sum_{n=1}^{N}|x_n|^{p}}$ represents the $\ell_{p}$‑norm ($p>0$).
We denote by $\bm{I}_{N}$ the $N \times N$ identity matrix.
The proximal operator of a function $\phi: \mathbb{R}^{N} \to \mathbb{R} \cup \{+ \infty\}$ is defined as $\prox_{\phi}(\bm{u}) \coloneq \argmin_{\bm{v} \in \mathbb{R}^{N}} \curbra{\frac{1}{2}\norm{\bm{u}-\bm{v}}^2_2+\phi(\bm{v})}$. 
The operators $\lceil \cdot \rceil$ and $\lfloor \cdot \rfloor$ denote the ceiling and floor functions, respectively.

\section{Modulo Sampling}

In typical sampling, the signal is clipped when the amplitude of the input signal exceeds the observable amplitude range of the ADC.
The clipped signal $f_{\text{cl}}(t)$ is obtained by limiting the amplitude of $f(t)$ to the range $[-\lambda, \lambda]$ ($\lambda > 0$), which is given by $f_{\text{cl}}(t) = \min(\lambda, \max(-\lambda, f(t)))$.

To avoid clipping, modulo sampling has been proposed~\cite{UnlimitedSamplingReconstruction-Bhandari-2021}.
In modulo sampling, the folded signal $f_\lambda(t)$ is obtained by applying the following operation to the input $f(t)$ as
\begin{align}
    f_\lambda(t) 
    = 
    M_\lambda(f(t))
    = 
    \{(f(t)+\lambda)\ \mathrm{mod}\ 2\lambda\}-\lambda.
\end{align}
The operation $M_\lambda(\cdot)$ folds the signal using the modulo operator when the amplitude of the input signal exceeds $[-\lambda, \lambda)$, and confines the signal amplitude within $[-\lambda, \lambda)$~\cite{ResidualRecoveryAlgorithm-Azar-2022}.
The sampled signal $f_\lambda[n]$ can be expressed as $f_\lambda[n] = f_\lambda(nT_s)$, where $T_s$ is the sampling period.

In modulo sampling, we need to reconstruct the original unfolded signal $f[n] = f(nT_s)$ from the folded samples $f_\lambda[n]$.
In the recovery of the original signal $f[n]$, we introduce the residual signal given by
\begin{align}
\begin{split}
  z[n] \coloneq f_\lambda[n]-f[n]
  \quad (\Leftrightarrow \, f_\lambda[n]=f[n]+z[n]),
\end{split}\label{eq:fold}
\end{align}
which is the difference between the folded signal $f_{\lambda}[n]$ and the original signal. 
Since $f_{\lambda}[n]$ is known, the problem of reconstructing $f[n]$ can be reduced to the reconstruction of the residual signal $z[n]$.

\section{Conventional ISTA-Based Residual Signal Reconstruction}
\label{sec:guidelines}

\subsection{Properties of Modulo Sampling}

\begin{figure}[tb]
  \centering
  \includegraphics[width=0.4\textwidth]{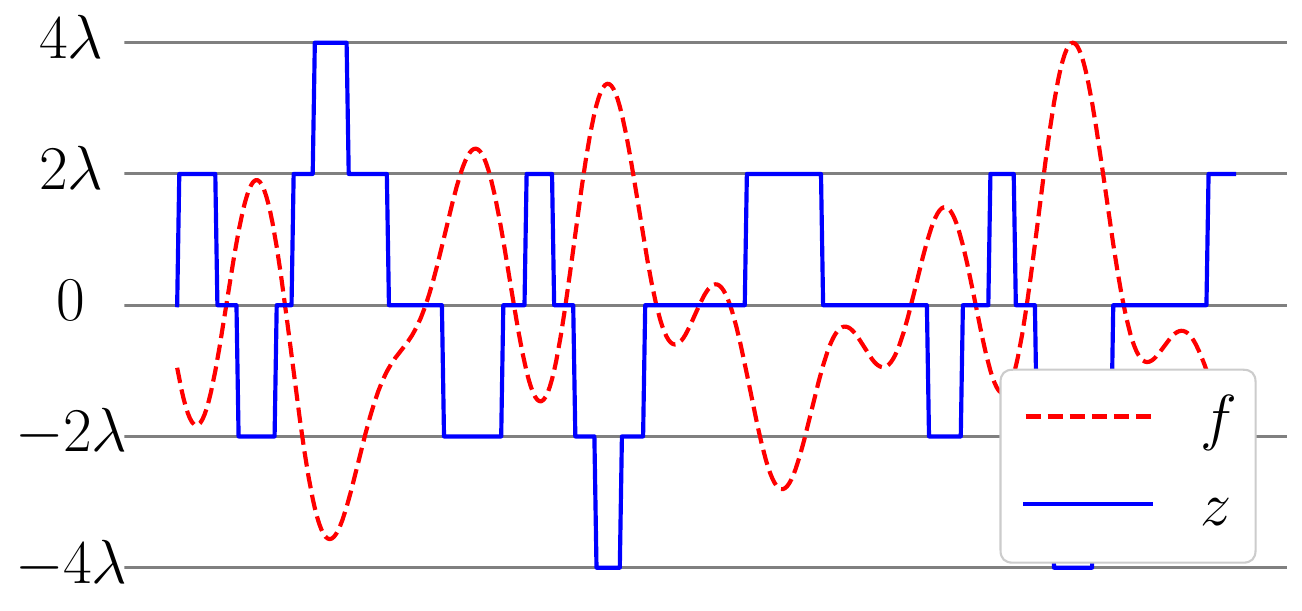}
  \caption{An original signal $f$ and its residual signal $z$}
  \label{fig:wave_z}
\end{figure}

As an effective method for recovering the original signal in modulo sampling, LASSO-$B^2R^2$ has been proposed \cite{LassoBasedFastResidual-Shah-2023}. 
This method uses the fact that the first-order difference of the residual signal $z[n]$ is sparse because $z[n]$ is typically piecewise-constant.
To see this, we show an example of the original signal $f[n]$ and the corresponding residual signal $z[n]$ in Fig.~\ref{fig:wave_z}. 
As can be seen from Fig.~\ref{fig:wave_z}, since $z[n]$ often takes the same value consecutively, its first-order difference $\hat{z}[n] \coloneq z[n]-z[n-1]$ becomes a sparse signal with many zero components.
Note that we here define $z[-1] = 0$ to ensure that the lengths of the original and the difference signals are identical.

LASSO-$B^2R^2$ also uses the properties of modulo sampling in the frequency domain.
The frequency spectrum of the original signal $f(t)$ is confined to the range $[-\omega_m, \omega_m]$, where $\omega_m$ represents the maximum angular frequency of the signal. 
On the other hand, the frequency spectrum of the folded signal and the residual signal extends beyond $[-\omega_m, \omega_m]$ as a result of the signal folding.
Thus, it is necessary to set the sampling frequency $\omega_s$ to be larger than twice the maximum frequency $\omega_m$ of the signal, i.e., the oversampling factor (OF) $\mathrm{OF}=\omega_s/(2\omega_m)$ is larger than one.

To analyze the original signal in the frequency domain, we first define discrete Fourier transform (DFT) of $f[n]$ as
\begin{align}
  F\paren{e^{j \frac{2 \pi k}{N}}} = \sum_{n=0}^{N-1}f[n]e^{-j \frac{2 \pi k n}{N}},
\end{align}
where $N$ is the signal length.
Given the relation $\omega T_{s} = 2 \pi k / N$ and $\omega_{s} = 2 \pi / T_{s}$, the condition for the frequency band where the original signal has no components, i.e., $\omega_m < |\omega| < \omega_s/2$, is equivalent to
\begin{align}
    \frac{\pi}{\mathrm{OF}} < \frac{2\pi k}{N} < 2\pi - \frac{\pi}{\mathrm{OF}}. \label{eq:k_condition}
\end{align}
For any integer $k$ satisfying~\eqref{eq:k_condition}, we have
\begin{align}
  F\paren{e^{j \frac{2 \pi k}{N}}} = 0. \label{eq:F_orig}
\end{align}

On the other hand, the frequency spectrum of the folded signal exists not only in the range $[-\omega_m, \omega_m]$ but also in the higher frequency bands owing to the signal folding effect.
Specifically, from~\eqref{eq:fold} and~\eqref{eq:F_orig}, DFT of the folded signal $F_\lambda(e^{j \frac{2 \pi k}{N}})$ satisfies
\begin{align}
  F_\lambda \paren{e^{j \frac{2 \pi k}{N}}} 
  = 
  Z\paren{e^{j \frac{2 \pi k}{N}}} \label{eq:z1}
\end{align}
for all integers $k$ satisfying~\eqref{eq:k_condition}.
Here, $F_\lambda \paren{e^{j \frac{2 \pi k}{N}}}$ and $Z\paren{e^{j \frac{2 \pi k}{N}}}$ are DFTs of the folded signal $f_\lambda[n]$ and the residual signal $z[n]$, respectively.

Let $\mathcal{K}$ be the set of integers $k$ that satisfy the condition $\frac{2\pi k}{N} \in (\frac{\pi}{\text{OF}},2\pi - \frac{\pi}{\text{OF}})$, and let $M = |\mathcal{K}|$ be the number of elements in this set. We consider a partial DFT matrix $\bm{V}\in \mathbb{C}^{M \times N}$ consisting only of rows corresponding to the elements $k$ in the set $\mathcal{K}$. In this matrix, the component specified by a certain $k \in \mathcal{K}$ and a column index $n$ is given by $v_{k,n}=e^{\frac{-j2\pi kn}{N}}$. Then, equation in~\eqref{eq:z1} can be written as
\begin{align}
  \bm{F}_\lambda=\bm{V}\bm{z}. \label{eq:z2}
\end{align}
Here, $\bm{F}_\lambda\in \mathbb{C}^{M}$ is a vector obtained by arranging the components of DFT of the folded signal $f_\lambda[n]$ corresponding to $\bm{V}$, and $\bm{z}\in \mathbb{R}^{N}$ is a vector obtained by arranging the residual signal $z[n]$.

The above discussion holds for the first-order difference of the signals as well.
We define the first-order difference $\hat{f}[n]$ of the signal $f[n]$ as $\hat{f}[n] := f[n] - f[n-1]$, as well as the first-order difference of the folded signal $f_\lambda[n]$ as $\hat{f_\lambda}[n] := f_\lambda[n] - f_\lambda[n-1]$. 
By taking the first-order difference of both sides of~\eqref{eq:fold}, we have $\hat{f_\lambda}[n]=\hat{f}[n]+\hat{z}[n]$.
At this time, $\hat{f}[n]$ is band-limited in the same way as $f[n]$, and hence we obtain
\begin{align}
  \hat{\bm{F}}_\lambda
  &=
  \bm{V}\hat{\bm{z}}. \label{eq:z4}
\end{align}

\subsection{Residual Signal Reconstruction Using ISTA} \label{sec:conventional}
In LASSO-$B^2R^2$, the first-order difference $\hat{\bm{z}}$ is reconstructed using~\eqref{eq:z4} and its sparsity. 
Since the components of $\hat{\bm{F}}_\lambda$ and $\bm{V}$ are complex numbers, we rewrite~\eqref{eq:z4} into a real-domain equation to construct a recovery problem in the real domain in this paper. 
Using the matrices formed by arranging the real and imaginary parts of $\hat{\bm{F}}_\lambda$ and $\bm{V}$ as $\hat{\bm{F}}_{\lambda}^{\mathrm{R}} = \sqb{ (\mathrm{Re}\ \hat{\bm{F}}_\lambda)^{\top} \ (\mathrm{Im}\ \hat{\bm{F}}_\lambda)^{\top}}^{\top}$ and $\bm{V}^{\mathrm{R}} = \sqb{ (\mathrm{Re}\ \bm{V})^{\top} \ (\mathrm{Im}\ \bm{V})^{\top}}^{\top}$, respectively, complex-valued model in~\eqref{eq:z4} can be rewritten as
\begin{align}
    \hat{\bm{F}}_{\lambda}^{\mathrm{R}}
  =
    \bm{V}^{\mathrm{R}}
  \hat{\bm{z}}. \label{eq:z5}
\end{align}

To recover $\hat{\bm{z}}$, LASSO-$B^2R^2$ solves the optimization problem based on LASSO \cite{RegressionShrinkageSelection-Tibshirani-1996} for sparse signal recovery as 
\begin{equation}
  \minimize_{\hat{\bm{z}} \in \mathbb{R}^{N}}\ 
  \frac{1}{2} \norm{\hat{\bm{F}}_{\lambda}^{\mathrm{R}}-\bm{V}^{\mathrm{R}}\hat{\bm{z}}}_{2}^{2}
  + \gamma \norm{\hat{\bm{z}}}_{1}. \label{eq:minl_1}
\end{equation}
The first term in~\eqref{eq:minl_1} is the difference between the observed value $\hat{\bm{F}}_{\lambda}^{\mathrm{R}}$ and $\bm{V}^{\mathrm{R}}\hat{\bm{z}}$, and the second term is a regularization term to exploit the sparsity of $\hat{\bm{z}}$. 
$\gamma$ ($> 0$) is a regularization parameter representing the weight of the regularization term.
The optimization problem in~\eqref{eq:minl_1} can be solved by the iterative shrinkage-thresholding algorithm (ISTA)~\cite{Daubechies2004-xa,SignalRecoveryProximal-Combettes-2005}.

After the reconstruction of $\hat{\bm{z}}$, $\bm{z}$ is obtained by taking the cumulative sum of $\hat{\bm{z}}$ as
\begin{align}
    z[n] = \sum_{k=0}^{n} \hat{z}[k] \quad \text{for } n=1, \dots, N. \label{eq:cumulative_sum}
\end{align}
Finally, the estimate of the original signal $f[n]$ is obtained from~\eqref{eq:fold}.

\section{Proposed Reconstruction Method} \label{sec:proposed}

\subsection{Fused Sparse Reconstruction (FSR)}

Conventional approaches such as LASSO-$B^2R^2$ suffer from two issues.  
First, because the reconstruction of the residual signal $\bm{z}$ involves the cumulative sum of the estimate of the first‑order difference $\hat{\bm{z}}$ as in~\eqref{eq:cumulative_sum}, a reconstruction error in a single element $z[n]$ can propagate and permanently contaminate subsequent estimate of the residual signal $\bm{z}$ and the original signal $\bm{f}$.  
To avoid this error propagation, it is preferable to reconstruct $\bm{z}$ directly, rather than reconstructing it through $\hat{\bm{z}}$.  
In this case, the observation model in~\eqref{eq:z2} can be used to reconstruct $\bm{z}$.  
To construct a real‑domain reconstruction algorithm, we consider the matrix $\bm{F}_\lambda^{\mathrm{R}} = \sqb{(\mathrm{Re}\ \bm{F}_\lambda)^{\top}\ (\mathrm{Im}\ \bm{F}_\lambda)^{\top}}^{\top}$ and rewrite~\eqref{eq:z2} as 
\begin{align}
    \bm{F}_\lambda^{\mathrm{R}}
  =
    \bm{V}^{\mathrm{R}}
  \bm{z}.
\end{align}
The task is then to reconstruct $\bm{z}$ from $\bm{F}_\lambda^{\mathrm{R}}$ and $\bm{V}^{\mathrm{R}}$.  
The second issue is that some conventional methods exploit only the sparsity of the first-order difference $\hat{\bm{z}}$ of the residual signal, although weak sparsity may exist in $\bm{z}$ itself. 
In practice, the residual signal $\bm{z}$ tends to be sparse when the original signal values are largely concentrated within $[-\lambda, \lambda)$, as $z[n] = 0$ whenever $f[n] \in [-\lambda, \lambda)$. 
This structural property suggests that incorporating a sparsity-promoting regularization on $\bm{z}$ itself can improve reconstruction performance.

To address both issues, we propose \emph{fused sparse reconstruction (FSR)}, which directly recovers $\bm{z}$ while leveraging the sparsity of both $\bm{z}$ and its first‑order difference.  
The proposed optimization problem for the reconstruction of $\bm{z}$ is formulated as  
\begin{align}
  \begin{split}
    \minimize_{\bm{z} \in \mathbb{R}^{N}}\ 
      \dfrac12 \bigl\lVert \bm{F}_\lambda^{\mathrm{R}}
      - \bm{V}^{\mathrm{R}}\bm{z} \bigr\rVert_{2}^{2}
      + \gamma_1 \lVert \bm{D}\bm{z} \rVert_{1}
      + \gamma_2 \lVert \bm{z} \rVert_{1}. 
  \end{split}\label{eq:proposed}
\end{align}
Here, $\bm{D}\in\mathbb{R}^{N\times N}$ is the first-order circular difference matrix, whose entries are $d_{i,j}=-1$ if $i=j$, $d_{i,j}=1$ if $i=(j-1)\bmod N$, and $d_{i,j}=0$ otherwise ($i,j\in{0,1,\dots,N-1}$).
The second term $\lVert \bm{D}\bm{z} \rVert_{1}$ in~\eqref{eq:proposed} enforces sparsity on the first‑order difference of $\bm{z}$, while the third term $\lVert \bm{z} \rVert_{1}$ promotes sparsity in $\bm{z}$ itself.  
The positive parameters $\gamma_1$ and $\gamma_2$ are the weights of these two regularizers, respectively.

\subsection{ADMM-Based Algorithm for FSR}
For the proposed optimization problem in~\eqref{eq:proposed}, we derive an algorithm based on ADMM~\cite{Eckstein1992-ra,DistributedOptimizationStatistical-Boyd-2010}.  
First, by introducing auxiliary variables $\bm{u}_1 \in \mathbb{R}^{N}$ and $\bm{u}_2 \in \mathbb{R}^{N}$, we rewrite~\eqref{eq:proposed} as  
\begin{align}
  \begin{split}
    &\minimize_{\bm{z} \in \mathbb{R}^{N}}\ 
         \dfrac12
         \lVert \bm{F}_\lambda^{\mathrm{R}}
         - \bm{V}^{\mathrm{R}}\bm{z} \rVert_{2}^{2}
         + \gamma_1 \lVert \bm{u}_1 \rVert_1
         + \gamma_2 \lVert \bm{u}_2 \rVert_1 \\[2pt]
    &\mathrm{subject\ to}\quad
       \bm{D}\bm{z} = \bm{u}_1,\;
       \bm{z}     = \bm{u}_2.
  \end{split} \label{eq:proposed_z_admm}
\end{align}
By letting $~\bm{u}=[\bm{u}^\top _1~ \bm{u}^\top _2]^{\top} \in \mathbb{R} ^{2N}$, 
~$\bm{\Phi} = [\bm{D}^{\top}~ \bm{I}_N]^\top\in \mathbb{R}^{2N \times N}$, $f(\bm{z})=\frac{1}{2}\norm{\bm{F}_\lambda^{\mathrm{R}}-\bm{V}^{\mathrm{R}}\bm{z}}_{2}^{2}$, and ~$g(\bm{u})= \gamma_1 \norm{\bm{u}_1}_1+\gamma_2 \norm{\bm{u}_2}_1$, 
we obtain 
\begin{align}
  \begin{split}
    &\minimize_{\bm{z} \in \mathbb{R}^{N},\;\bm{u} \in \mathbb{R}^{2N}}\ 
       f(\bm{z}) + g(\bm{u}) \\
    &\quad \mathrm{subject\ to}\quad \bm{\Phi}\bm{z} = \bm{u}.
  \end{split}\label{eq:opt_ADMM}
\end{align}
The ADMM iterations for~\eqref{eq:opt_ADMM} are given by 
\begin{align}
  \bm{z}^{(i+1)}&=\argmin_{\bm{z}\in\mathbb{R}^{N}}
        \Bigl\{f(\bm{z})+\dfrac{\rho}{2}
          \|\bm{\Phi}\bm{z}-\bm{u}^{(i)}+\bm{y}^{(i)}\|_{2}^{2}\Bigr\},
          \label{eq:z_admm}\\
  \bm{u}^{(i+1)}&=\prox_{\frac{1}{\rho}g}
        (\bm{\Phi}\bm{z}^{(i+1)}+\bm{y}^{(i)}),\label{eq:u_admm}\\
  \bm{y}^{(i+1)}&=\bm{y}^{(i)}+\bm{\Phi}\bm{z}^{(i+1)}-\bm{u}^{(i+1)},
          \label{eq:y_admm}
\end{align}
where $\rho$ ($>0$) is the parameter, $i$ is the iteration index, and $\bm{y}^{(i)}\in\mathbb{R}^{2N}$ is the scaled dual variable.

Both updates in~\eqref{eq:z_admm} and~\eqref{eq:u_admm} have explicit expressions.  
Solving~\eqref{eq:z_admm} yields
\begin{align}
  \bm{z}^{(i+1)}
  &=
  \bigl(\rho(\bm{D}^{\top}\bm{D}+\bm{I}_N)
        +{\bm{V}^{\mathrm{R}}}^{\!\top}\bm{V}^{\mathrm{R}}\bigr)^{-1}
        \notag\\
  &\quad\cdot
  \bigl(\rho\bm{D}^{\top}(\bm{u}_1^{(i)}-\bm{y}_1^{(i)})
        +\rho(\bm{u}_2^{(i)}-\bm{y}_2^{(i)})
        +{\bm{V}^{\mathrm{R}}}^{\!\top}\bm{F}_\lambda^{\mathrm{R}}\bigr).
        \label{eq:z_update}
\end{align}
Here, $\bm{y}_1 \in \mathbb{R}^{N}$ and $\bm{y}_2 \in \mathbb{R} ^{N}$ are variables obtained by partitioning $\bm{y}$ such that $\bm{y}=[\bm{y}^\top _1~ \bm{y}^\top _2]^{\top}$, in the same manner as $\bm{u}$.
The update in~\eqref{eq:u_admm} decouples as
\begin{align}
  \bm{u}^{(i+1)}=\begin{bmatrix}
    \prox_{\frac{\gamma_1}{\rho}\|\cdot\|_1}
       (\bm{D}\bm{z}^{(i+1)}+\bm{y}^{(i)}_1)\\
    \prox_{\frac{\gamma_2}{\rho}\|\cdot\|_1}
       (\bm{z}^{(i+1)}+\bm{y}^{(i)}_2)
  \end{bmatrix}.\label{eq:u_admm3}
\end{align}
The proximal operator for the $\ell_1$ norm becomes an element-wise soft-thresholding function given by $\prox_{\tau \|\cdot\|_1}(x) = \sign (x)\max(\abs{x}-\tau,0)$, where $\sign(\cdot)$ is the sign function. 

After we obtain the estimate of $\bm{z}$, we apply a rounding operation to enforce the estimate to be an integer multiple of $2\lambda$.
The proposed reconstruction algorithm is presented in Algorithm~\ref{alg:SSR_ADMM}.

\begin{algorithm}[tb]
  \caption{Proposed Fused Sparse Reconstruction (FSR)}
  \label{alg:SSR_ADMM}
  \begin{algorithmic}[1]
      \REQUIRE $f_\lambda[n]\;(0\le n<N),\;\lambda,\;I,\;\mathrm{OF},\;N,\;\bm{V}^{\mathrm{R}}$
      \ENSURE $f[n]$
      \STATE \textbf{Initialization:} $\gamma_{1}, \gamma_{2}, \rho>0$ and $\hat{\bm{z}}^{(0)}\sim\mathcal{N}(0,1)$
      \STATE Compute $F_\lambda (e^{j \frac{2 \pi k}{N}}),\:\forall k\in \mathbb{Z}~\text{and}~\frac{2\pi k}{N} \in (\frac{\pi}{\text{OF}},2\pi - \frac{\pi}{\text{OF}})$
      \FOR{$i = 0$ \textbf{to} $I-1$}
          \STATE Compute $\bm{z}^{(i+1)}$ via~\eqref{eq:z_update}
          \STATE $\bm{u}^{(i+1)}_1 \leftarrow
                 \prox_{\frac{\gamma_1}{\rho}\lVert\cdot\rVert_1}
                 \bigl(\bm{D}\bm{z}^{(i+1)}+\bm{y}^{(i)}_1\bigr)$
          \STATE $\bm{u}^{(i+1)}_2 \leftarrow
                 \prox_{\frac{\gamma_2}{\rho}\lVert\cdot\rVert_1}
                 \bigl(\bm{z}^{(i+1)}+\bm{y}^{(i)}_2\bigr)$
          \STATE $\bm{y}_{1}^{(i+1)} \leftarrow
                 \bm{y}^{(i)}_1
                 + \bm{D}\bm{z}^{(i+1)}
                 - \bm{u}_{1}^{(i+1)}$
          \STATE $\bm{y}_{2}^{(i+1)} \leftarrow
                 \bm{y}_{2}^{(i)}
                 + \bm{z}^{(i+1)}
                 - \bm{u}_{2}^{(i+1)}$
      \ENDFOR
      \STATE $\bm{z} \leftarrow \bm{z}^{(I)}$
      \STATE $\bm{z} \leftarrow
              \bigl\lceil
                \tfrac{\lfloor \bm{z}/\lambda \rfloor}{2}
              \bigr\rceil
              \cdot 2\lambda$
      \STATE $f[n] \leftarrow f_\lambda[n] - z[n]$
  \end{algorithmic}
\end{algorithm}
The computational cost of Algorithm~\ref{alg:SSR_ADMM} is dominated by the update step for $\bm{z}$ in~\eqref{eq:z_update}. 
A straightforward implementation of this step would require a matrix inversion, resulting in a computational cost of $O(N^3)$. 
However, since the inverse matrix admits the closed-form expression 
$(\rho(\bm{D}^{\top}\bm{D}+\bm{I}_N) + {\bm{V}^{\mathrm{R}}}^{\top}\bm{V}^{\mathrm{R}})^{-1} = \bm{F}^{\mathsf{H}} ( \rho(\bm{\Lambda}_D + \bm{I}) + \bm{\Lambda}_V )^{-1} \bm{F}$ with the inverse of a diagonal matrix, the computational complexity per iteration becomes $O(N \log N)$. 
Detailed derivations are given in Appendix. 

\section{Simulation Results}

We evaluate the performance of the proposed method via computer simulations. 
A test signal with maximum frequency $1/T_s \times 1/\text{OF}$ is first generated by superposing five random sine waves whose amplitudes, frequencies, and phases drawn independently from uniform distributions. 
This composite waveform is then normalized so that its maximum amplitude becomes~$1$.
Modulo sampling with width~$\lambda = 0.25$ is applied to this signal.  
Zero‑mean Gaussian noise generated according to each specified signal‑to‑noise ratio (SNR) is added to the signal.  
For each noisy signal, the reconstruction is performed $250$ times, and the averaged NMSE between the original and reconstructed signals given by $\mathrm{NMSE} = \|\bm{f}-\bm{f}_{\mathrm{est}}\|_{2}^{2} / \|\bm{f}\|_{2}^{2}$ is evaluated, where $\bm{f} = \sqb{f[0], \dotsc, f[N-1]}^{\top} \in \mathbb{R}^{N}$ denotes the original signal and $\bm{f}_{\mathrm{est}}$ is the corresponding reconstructed signal.
The parameters are set to $N=1024$, $T_s=0.01$, $I=150$, $\gamma_1=1$, $\gamma_{2}=0.01$ and $\rho = 2.0$.

\begin{figure}[tb]
  \centering
  \includegraphics[width=0.4\textwidth]{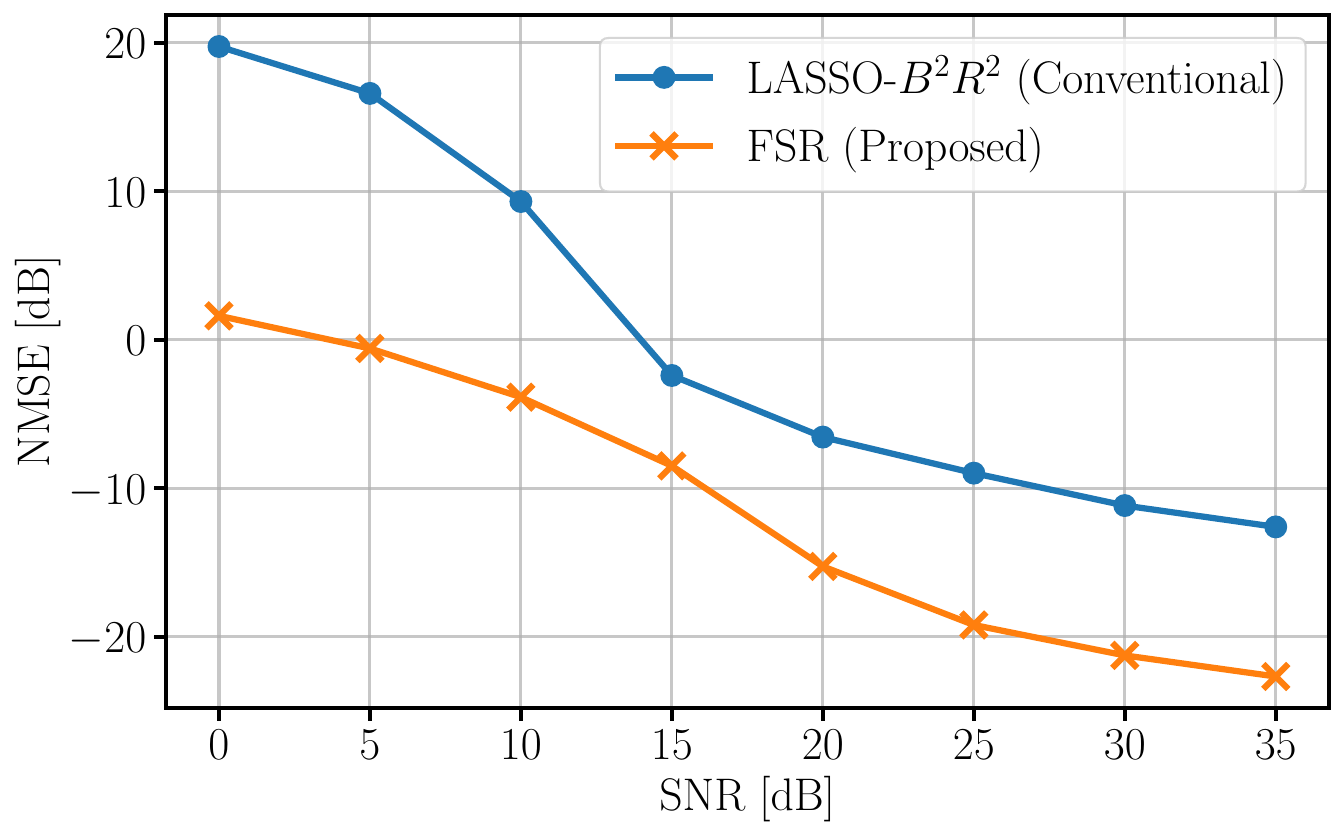}
  \caption{Reconstruction results for $\text{OF}=6$.}
  \label{fig:res2}
\end{figure}

\begin{figure}[tb]
  \centering
  \includegraphics[width=0.4\textwidth]{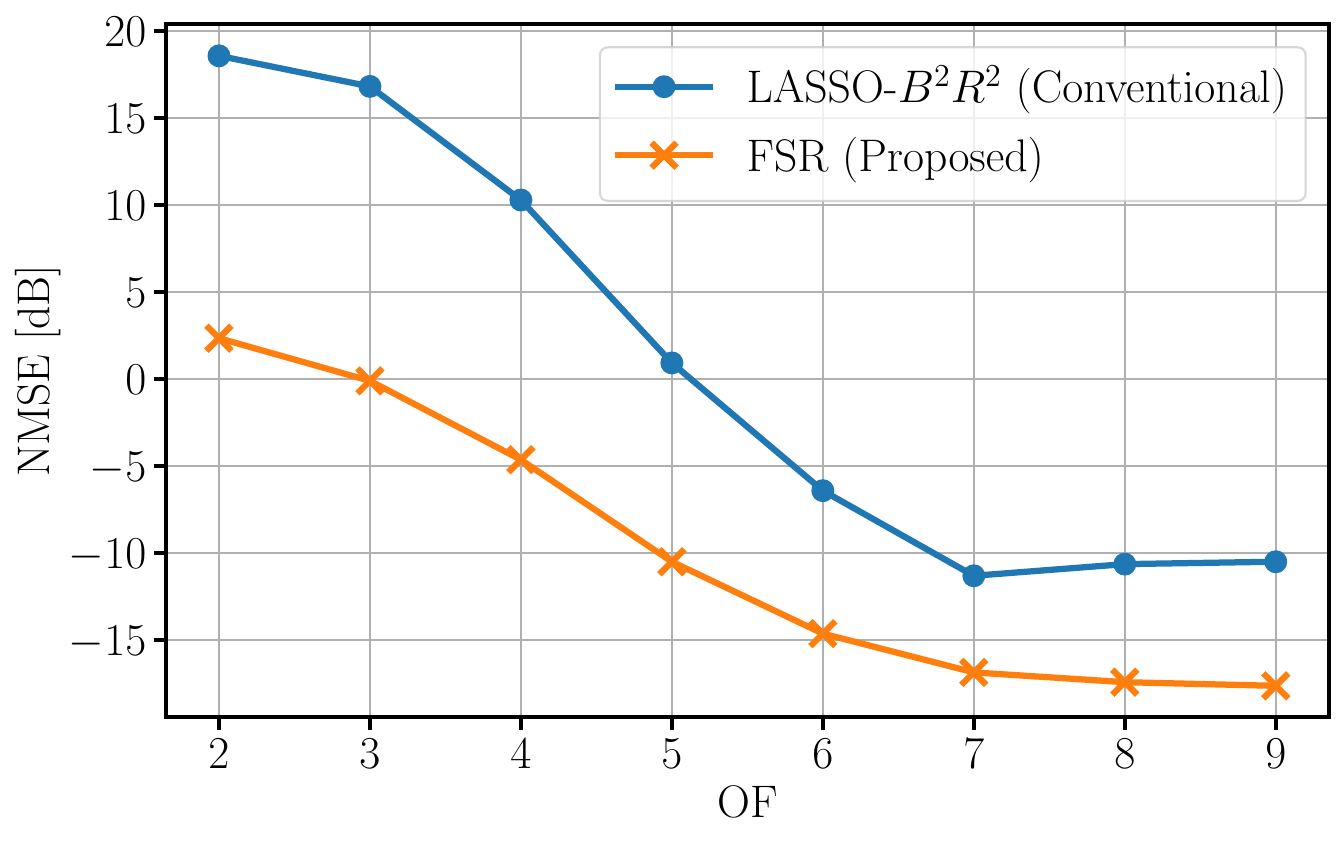}
  \caption{Reconstruction results for $\text{SNR}=20$~dB.}
  \label{fig:res3}
\end{figure}

Fig.~\ref{fig:res2} shows the NMSE performance when OF is $6$ and SNR is varied from $0$ to $35$~dB.  
Fig.~\ref{fig:res3} plots the NMSE performance when SNR is $20$~dB and OF is varied from $2$ to $9$.
From the figures, we can see that the proposed method consistently achieves a smaller NMSE than LASSO-$B^{2}R^{2}$, which achieves better performance than several other conventional methods~\cite{LassoBasedFastResidual-Shah-2023}.  
This improvement is attributed to the promotion of sparsity in~$\bm{z}$ by the new regularization term and to the elimination of error propagation achieved by directly recovering~$\bm{z}$.

\section{Conclusion}

In this study, we have proposed an optimization problem that directly reconstructs the residual signal for signal reconstruction in modulo sampling.  
The proposed formulation exploits both the high‑frequency characteristics of the sampled data and two types of sparsity: in the residual signal itself and in its first‑order difference. 
For the proposed optimization problem, we have derived an efficient ADMM‑based algorithm.  
Computer simulations have demonstrated that the proposed method achieves superior reconstruction accuracy over the conventional LASSO-$B^{2}R^{2}$.

\appendix

\section{Efficient Computation of Inverse Matrix\\in the Proposed FSR}
This appendix describes the efficient computation of the inverse matrix of 
\begin{align}
    \bm{A} \coloneq \rho(\bm{D}^{\top}\bm{D} + \bm{I}) + {\bm{V}^{\mathrm{R}}}^{\top}\bm{V}^{\mathrm{R}}
\end{align} 
used in the $\bm{z}$-update step of the proposed algorithm. 
We will show that we can efficiently multiply $\bm{A}^{-1}$ to a vector $\bm{b}$ (i.e., compute $\bm{x}=\bm{A}^{-1}\bm{b}$) with $\mathcal{O}(N \log N)$ complexity by exploiting the diagonalization properties of the constituent matrices. 

\subsection{Diagonalization of the Difference Matrix}
Assuming periodic boundary conditions, the difference matrix $\bm{D}$ is a circulant matrix, and thus $\bm{D}^{\top}\bm{D}$ is also circulant. 
It can be diagonalized by the normalized discrete Fourier transform (DFT) matrix $\bm{F} \in \mathbb{C}^{N \times N}$ as $\bm{D}^{\top}\bm{D} = \bm{F}^{\mathsf{H}} \bm{\Lambda}_D \bm{F}$, where $(\cdot)^{\mathsf{H}}$ denotes the Hermitian transpose. 
The $k$-th diagonal element of $\bm{\Lambda}_D$ is given by
\begin{align}
    \lambda_{D,k} = \left| 1 - e^{-j \frac{2\pi k}{N}} \right|^2 = 4 \sin^2\left(\frac{\pi k}{N}\right).
\end{align}

\subsection{Diagonalization of the Observation Matrix}
To diagonalize the second term ${\bm{V}^{\mathrm{R}}}^{\top}\bm{V}^{\mathrm{R}}$ using the DFT matrix, we first establish the relationship between ${\bm{V}^{\mathrm{R}}}$ and the complex observation matrix $\bm{V} = \mathrm{Re}\ \bm{V} + j \, \mathrm{Im}\ \bm{V} \in \mathbb{C}^{M \times N}$. The real-valued observation matrix $\bm{V}^{\mathrm{R}}$ is defined by stacking the real and imaginary parts as
\begin{align}
    \bm{V}^{\mathrm{R}} \coloneq \begin{bmatrix} \mathrm{Re}\ \bm{V} \\ \mathrm{Im}\ \bm{V} \end{bmatrix} \in \mathbb{R}^{2M \times N}.
\end{align}
A key identity relates this real Gram matrix to the complex one:
\begin{align}
    {\bm{V}^{\mathrm{R}}}^{\top}\bm{V}^{\mathrm{R}} 
    &= 
    (\mathrm{Re}\ \bm{V})^{\top} (\mathrm{Re}\ \bm{V}) 
    + (\mathrm{Im}\ \bm{V})^{\top} (\mathrm{Im}\ \bm{V}) \\
    &= 
    \mathrm{Re} (\bm{V}^{\mathsf{H}}\bm{V}).
    \label{eq:appendix_identity}
\end{align}

Now, consider the case of partial Fourier sensing, where $\bm{V} = \bm{S}\bm{F}$ with a row-selection matrix $\bm{S}$. In this case, $\bm{V}^{\mathsf{H}}\bm{V} = \bm{F}^{\mathsf{H}} \bm{\Lambda}_{\mathrm{mask}} \bm{F}$, where $\bm{\Lambda}_{\mathrm{mask}} = \bm{S}^\top\bm{S}$ is a diagonal sampling mask. 
Substituting this into \eqref{eq:appendix_identity} yields
\begin{align}
    {\bm{V}^{\mathrm{R}}}^{\top}\bm{V}^{\mathrm{R}} &= \mathrm{Re}(\bm{F}^{\mathsf{H}} \bm{\Lambda}_{\mathrm{mask}} \bm{F}) \notag \\
    &= \frac{1}{2}\left(\bm{F}^{\mathsf{H}} \bm{\Lambda}_{\mathrm{mask}} \bm{F} + (\bm{F}^{\mathsf{H}} \bm{\Lambda}_{\mathrm{mask}} \bm{F})^*\right),
\end{align}
where $(\cdot)^{*}$ denotes the complex conjugate. 

Let $\bm{\Pi}\in\mathbb{R}^{N\times N}$ be the frequency-reversal permutation matrix defined by
\begin{align}
[\bm{\Pi}]_{k,\ell}=
\begin{cases}
1, & \ell = (-k)_N,\\
0, & \text{otherwise},
\end{cases}
\qquad (-k)_N\coloneq(-k)\bmod N.
\end{align}
Then $\bm{\Pi}^{\top}=\bm{\Pi}^{-1}=\bm{\Pi}$.
For the normalized DFT matrix $[\bm{F}]_{m,n}=\frac{1}{\sqrt{N}}e^{-j\frac{2\pi mn}{N}}$, we have
\begin{align}
[\bm{\Pi}\bm{F}]_{m,n}= [\bm{F}]_{(-m)_N,n}
=\frac{1}{\sqrt{N}}e^{-j\frac{2\pi (-m)n}{N}}
=[\bm{F}^*]_{m,n},
\end{align}
hence $\bm{\Pi}\bm{F}=\bm{F}^*$ and therefore $\bm{F}^{\top}=(\bm{F}^*)^{\mathsf H}=(\bm{\Pi}\bm{F})^{\mathsf H}=\bm{F}^{\mathsf H}\bm{\Pi}$.

Since $\bm{\Lambda}_{\mathrm{mask}}$ is real diagonal, we have 
\begin{align}
(\bm{F}^{\mathsf{H}}\bm{\Lambda}_{\mathrm{mask}}\bm{F})^*
&= (\bm{F}^{\mathsf{H}})^*\,\bm{\Lambda}_{\mathrm{mask}}^*\,\bm{F}^* \notag\\
&= \bm{F}^\top\,\bm{\Lambda}_{\mathrm{mask}}\,\bm{F}^* \notag\\
&= (\bm{F}^{\mathsf{H}}\bm{\Pi})\,\bm{\Lambda}_{\mathrm{mask}}\,(\bm{\Pi}\bm{F}) \notag\\
&= \bm{F}^{\mathsf{H}}(\bm{\Pi}\bm{\Lambda}_{\mathrm{mask}}\bm{\Pi})\bm{F}.
\end{align}
Thus, the observation term is diagonalized by the DFT matrix as
\begin{align}
    {\bm{V}^{\mathrm{R}}}^{\top}\bm{V}^{\mathrm{R}} = \bm{F}^{\mathsf{H}} \bm{\Lambda}_V \bm{F},
\end{align}
where the diagonal matrix $\bm{\Lambda}_V$ is defined as
\begin{align}
    \bm{\Lambda}_V \coloneq \frac{\bm{\Lambda}_{\mathrm{mask}} + \bm{\Pi} \bm{\Lambda}_{\mathrm{mask}} \bm{\Pi}}{2}.
\end{align}
The $k$-th diagonal element $\lambda_{V,k}$ of $\bm{\Lambda}_{V}$ is therefore given by the average of the mask at symmetric frequencies as
\begin{align}
    \lambda_{V,k} = \frac{\lambda_{\mathrm{mask},k} + \lambda_{\mathrm{mask},(-k)_N}}{2}, 
\end{align}
where $\lambda_{\mathrm{mask},k}$ is the $k$-th diagonal element of $\bm{\Lambda}_{\mathrm{mask}}$. 
\subsection{Summary of Fast Inversion}
Since both $\bm{D}^{\top}\bm{D}$ and ${\bm{V}^{\mathrm{R}}}^{\top}\bm{V}^{\mathrm{R}}$ are simultaniously diagonalized by $\bm{F}$, the matrix $\bm{A}$ is expressed as
\begin{align}
    \bm{A} 
    &= 
    \bm{F}^{\mathsf{H}} \left( \rho (\bm{\Lambda}_D + \bm{I}) + \bm{\Lambda}_V \right) \bm{F} \\
    &= 
    \bm{F}^{\mathsf{H}} \bm{\Lambda}_A \bm{F}, \label{eq:A_eigendecomp}
\end{align}
where $\bm{\Lambda}_A = \rho (\bm{\Lambda}_D + \bm{I}) + \bm{\Lambda}_V$ is the diagonal matrix. 
From~\eqref{eq:A_eigendecomp}, we have $\bm{A}^{-1}=\bm{F}^{\mathsf{H}}\bm{\Lambda}_A^{-1}\bm{F}$. 
Thus, multiplication by $\bm{A}^{-1}$ can be computed via FFT, element-wise division by the diagonal entries of $\bm{\Lambda}_A$, and IFFT. 
The overall computational complexity is reduced to $\mathcal{O}(N\log N)$ from $\mathcal{O}(N^{3})$ for the straightforward implementation.

\bibliographystyle{IEEEtran}
\bibliography{SPL}

\end{document}